\documentclass[11pt]{article}
\usepackage{mathtext}
\usepackage[T2A]{fontenc}
\usepackage{amsfonts}
\usepackage[intlimits]{amsmath}
\usepackage{amsthm}
\usepackage{amssymb}
\usepackage{setspace}
\usepackage{graphicx}

\newcommand{\p}{\partial}
\newcommand{\be}{\begin{eqnarray}}
\newcommand{\ee}{\end{eqnarray}}
\newcommand{\de}{\delta}

\newcommand{\eps}{\varepsilon}

\newcommand{\R}{\mathcal{R}}
\newcommand{\vecc}[1]{{\rm\bf{#1}}}

\newcommand{\ba}[1]{\begin{array}{#1}}
\newcommand{\ea}{\end{array}}

\newcommand{\da}{\dot{a}}

\title{
\begin{flushright}
 \mbox{\normalsize ITEP/TH-30/10}
\end{flushright}
What do we learn from the CMB observations?}

\date{}

\author{V. A. Rubakov$^1$, A. D. Vlasov$^2$ \\
$^1$ Institute for nuclear research of the Russian Academy of sciences, Moscow; \\
$^1$ email: rubakov@ms2.inr.ac.ru \\
$^2$ Institute for theoretical and experimental physics, Moscow;\\
$^2$ email: vlasov.ad@gmail.com
}

\begin{document}

\maketitle

\begin{abstract}
 We give an account, at nonexpert and quantitative level, of physics behind the CMB temperature anisotropy and polarization and their peculiar features. We discuss, in particular, how cosmological parameters are determined from the CMB measurements and their combinations with other observations. We emphasize that CMB is the major source of information on the primordial density perturbations and, possibly, gravitational waves, and discuss the implication for our understanding of the extremely early Universe.
\end{abstract}

\tableofcontents

\section{Introduction}

\hspace*{\parindent}The existence of cosmic microwave background radiation (CMB) is one of the keystones of cosmology. It was discovered in 1964 by Pensias and Wilson, and it earned them  the Nobel prize.
CMB  comes from the epoch when the Universe was much hotter and much younger than today. The approximate age of the Universe at the time CMB was released equals $380$ thousand years, while the current age of the Universe is about 14 billion years. Presently, the radiation has the temperature $T=2.725 K$. Among all signals we receive directly, CMB radiation is believed to be one of the youngest\footnote{Earlier epoch of Big Bang Nucleosynthesis (starting from about 1 second after Big Bang) is probed by measuring light element abundances.}. 

\begin{figure}[h!]
\begin{center}
\includegraphics[width=0.8\linewidth]{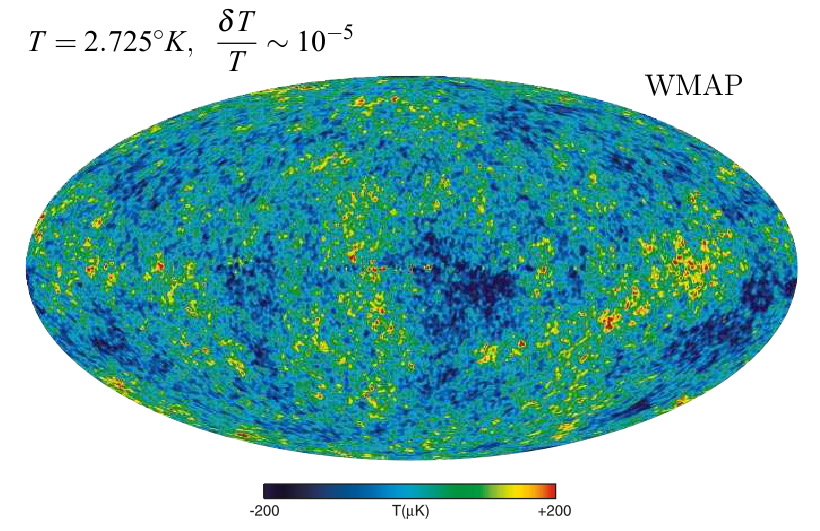}
\caption{~The CMB temperature map, obtained by WMAP experiment \cite{WMAP}.
The brighter a region is, the hotter radiation comes from it.}
\label{anisotropy}
\end{center}
\end{figure}

Later, the anisotropy of CMB was discovered. This anisotropy is very small. The largest is the dipole component of order $\de T/T\approx 10^{-3}$. It is believed to arise due to the motion of the Solar system with respect to CMB. All other multipoles are much lower, the maximal ones are of 
order $\de T/T \sim 10^{-4} - 10^{-5}$ (see Fig.\ref{anisotropy}).

 It turns out that one can determine many parameters of our Universe by studying  this anisotropy. CMB anisotropy is caused both by inhomogeneities in the early Universe and by the effects of propagation of CMB photons through the Universe at later time.  An example of the latter is the Sunyaev--Zel'dovich effect, it is due to scattering of CMB photons off free electrons in interstellar plasma. This effect distorts CMB angular spectrum at small angular scales. There are other effects with similar consequences. For 
example, recombination is not an instantaneous process, 
the surface of last scattering has finite width, and
this smooths out the
CMB angular spectrum at small angular scales. Also, just before recombination the mean free path of photons drastically increases. The photons  travel freely for rather long time, so they diffuse away from over-dense regions to regions of lower density, and this also smears out small inhomogeneities. This effect is called  Silk damping. Because of 
these effects, the CMB anisotropy spectrum is not 
particularly informative from the cosmological point of view at angular scales of a few arc minutes
and smaller. 

Last scattering of the CMB photons occurs almost at the same time as the recombination does, since the main process of interaction of photons with matter in this case is the Thomson scattering off free electrons.
%  The temperature of recombination depends only on the density of the media only logarithmically. So, the dependence on the cosmological parameters, such as Hubble parameter, $\Omega$'s, etc. is weak.
The temperature of last scattering depends on the cosmological parameters very weakly (weaker than logarithmically), it is mainly determined by the potential of ionization of hydrogen and other world constants. The up-to-date values of the temperature and redshift of last scattering are:
\be
T_r=2970 \, {\rm K}=0.26~{\rm eV}, \\
z_r=1090. 
\ee
At the time of last scattering, the Universe was in matter--dominated regime,  i. e., the contribution of mass density of nonrelativistic matter to the total energy density  is the largest, or, in other words, the overall pressure is much smaller than the overall energy density: $p_{\rm tot}\ll \rho_{\rm tot}$. 

CMB gives us the photographic picture of the Universe at the time of last
scattering (modulo aforementioned effects). Hence, the CMB anisotropy
reflects the fact that the Universe was slightly inhomogeneous at that time.
 The inhomogeneities in the following components of cosmic plasma contribute to the CMB anisotropy:
\begin{enumerate}
\item Baryon--electron--photon plasma.
\item Dark matter.
\item Neutrinos.
\item Gravitational waves.
\end{enumerate}
Rather complex structure of the CMB angular spectrum is due to the
interplay between the effects caused by the first two of these
components. 

The purpose of this review is to discuss what information about our Universe
one extracts from the CMB data. There are two classes of properties one is
interested in:
\begin{enumerate}
\item Properties of primordial perturbations which, as we will see,
were buil before the hot stage of the cosmological evolution.
This is a key for revealing the origin of primordial perturbations and,
ultimately, understanding
the epoch preceding the hot stage. The existing data
are consistent with the picture that primordial density perturbations
and gravitational potentials induced by them had rather simple properties,
consistent with the inflationary mechanism of their generation,
and that there were no primordial gravitational waves. As we discuss in
this review, future CMB 
measurements are sensitive to primordial gravity waves
at the level predicted by simple inflationary models. The observation of
the effects induced in CMB by primordial gravity waves would be an
outstanding discovery.
\item Properties of the Universe shortly before recombination and
later, up until the present epoch. These include the energy balance
(photons, neutrinos, baryons, dark matter, dark energy), spatial curvature,
etc. 
\end{enumerate}

Studying the CMB temperature anisotropy and polarization is 
sufficient by itself to measure some of these properties. There are 
numerous cases, however, when the CMB data alone are insufficient for
determining some of the cosmological parameters, because these data are
approximately degenerate in the parameters (i.e., they depend on certain
combinations of the parameters). The degeneracy is lifted by invoking
other cosmological data, as we illustrate in this review.

There are good books and reviews on CMB, see, for example,
\cite{Peacock,Dodelson,Mukhanov,Naselsky,Weinberg,Giovannini,Durrer,GorbunovRubakov,Kosowsky,Challinor}.
% Refs.~[2-11].
The reader is invited to consult them, in particular, for
references to original works. In this review we mostly refer to
papers containing concrete results we make use of.
\section{Observables}
\begin{figure}[h!]
\begin{center}
\includegraphics[width=0.8\linewidth]{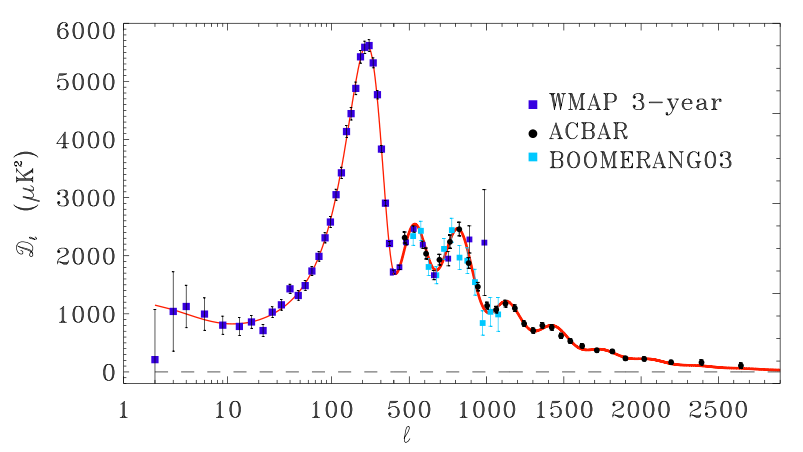}
\caption{~The angular spectrum of the CMB temperature anisotropy  \cite{ACBAR}. The line is a prediction of the standard $\Lambda CDM$ model. The quantity in vertical axis  is $D_l$ defined by (\ref{defdl}).}
\label{anisotropyspectrum}
\end{center}
\end{figure}

%Any function on the sphere, and particularly
The temperature variation on celestial sphere, i. e., the function 
$\de T(\phi,\theta) \equiv \de T(\rm{\bf{n}})$, 
can be expanded in spherical harmonics:
\be
\de T(\vecc{n})=\sum_{lm}a_{lm}Y_{lm}(\theta,\phi).
\ee
All observations today support the hypothesis that $a_{lm}$ {\it are independent  Gaussian random variables}. Gaussianity means that
\be
P(a_{lm})da_{lm}=\frac{1}{\sqrt{2\pi C_l}} e^{-\frac{a_{lm}^2}{2C_l}}da_{lm},
\ee
where $P(a_{lm})$ is the probability density for the random variable
$a_{lm}$. For a hypothetical ensemble of Universes like ours, the mean values
of products of the coefficients $a_{lm}$ would obey
\be
\langle a_{lm} a_{l^\prime m^\prime} \rangle = C_l \delta_{l l^\prime}
\delta_{m m^\prime}.
\label{may29-6}
\ee
We can measure only one Universe, but this formula is  still used to
extract the angular spectrum $C_l$ from the data. For given $l$, there
are $(2l+1)$ independent coefficients $a_{lm}$, so there exists
an irreducible statistical uncertainty of order $\delta C_l/C_l \sim
1/\sqrt{2l+1}$, called cosmic variance. It is particularly pronounced
at small $l$ and, indeed, it is much larger than the experimental
errors in this part of the angular spectrum (as an example,
error bars in the left part of Fig.~\ref{anisotropyspectrum} are
precisely due to the cosmic variance).

In the case of Gaussian random variables, all odd correlators are zero and all even correlators are expressed through the two-point correlator 
according to the Wick theorem. Under this distribution, the mean value of
the
temperature fluctuation squared equals:
\be
 \langle[\de T(\vecc{n})]^2\rangle=\sum_l\frac{2l+1}{4\pi}C_l\approx \int \frac{dl}{l}\frac{l(l+1)}{2\pi}C_l=\int \frac{dl}{l} D_l. \label{defdl}
\ee
Usually, graphs presenting the anisotropy spectrum show $D_l$ (for example, in Fig.\ref{anisotropyspectrum}). It is worth noting that the approximate
relationship between the multipole number $l$ and angular scale
is $\Delta \theta \simeq \pi/l$. Hence, the first peak
in  Fig.\ref{anisotropyspectrum} corresponds to temperature fluctuation of
angular size of about $1^0$.

Gaussianity of temperature fluctuations is a very interesting property.
 The linear evolution preserves Gaussianity, so this property means that the  
primordial density perturbations are Gaussian random field. 
We recall in this regard that 
vacuum fluctuations of free quantum fields are also Gaussian. Hence, we have an observational hint that 
{\it the mechanism of the generation of
primordial perturbations 
is the amplification of vacuum fluctuations of some linear quantum field(s)}.
% It is this what occurs in inflationary paradigm, but also in some other models.
This is indeed the case in the inflationary scenario, and also in some other models.
\section{Density perturbations and CMB temperature anisotropy}
\subsection{Background metric}
To set the stage, let us begin with background. The background metric 
is the metric of homogeneous and isotropic space--time (Friedmann--Robertson--Walker metric):
\be
ds^2=dt^2-a(t)^2\left[\frac{dr^2}{1-\kappa r^2}+r^2(d\theta^2+\sin^2\theta d\phi^2)\right], \label{friedman}
\ee
where $a(t)$ is the scale factor, $\kappa$ is the spatial curvature parameter. $\kappa=0,+1,-1$ correspond to the Euclidean 3--space, 3--sphere and 3--hyperboloid, respectively. In most part of this review we consider the case $\kappa=0$ and use the equivalent form of the spatially flat metrics:
\be
 ds^2=dt^2-a(t)^2(dx_1^2+dx_2^2+dx_3^2).
\ee
 If we neglect the possible peculiar motion of cosmological observers, each observer has definite time-independent $x$-coordinates. Note that $x_1^2+x_2^2+x_3^2$ is not a square of physical distance even in the spatially flat Universe ($\kappa=0$). The physical distance squared is $a^2(x_1^2+x_2^2+x_3^2)$, and this corresponds to the fact that the physical distance between two cosmological observers grows in time as $a(t)$ increases. This is precisely the expansion of the Universe. The coordinates $x$ are called comoving coordinates. 

In general, the dynamics of the scale factor is determined by
the Friedmann equation, which reads:
\be
\left(\frac{\da}{a}\right)^2=\frac{8\pi G}{3}\rho - \frac{\kappa}{a^2},
\label{jun1-1}
\ee
where $\rho$ is the total energy density in the Universe.
The energy balance equation (the relation $dE=-pdV$ applied to
the comoving volume $a^3$) is
\be
  d\rho = - 3(p+\rho) \frac{da}{a},
\ee
where $p$ is pressure.
For nonrelativistic 
matter (dark matter, baryons and also massive neutrinos at late epoch)
one has $p \ll \rho$, and the energy density decreases as 
$1/a^3$ (mass density gets diluted in inverse proportion to
the comoving volume). 
For
ultrarelativistic matter, for example, photons and neutrinos
at early epoch, 
the equation of state is $p=\rho/3$ and 
$\rho \propto 1/a^4$ (energy density decreases also because photons get
redshifted). Equation (\ref{jun1-1}) then implies that
if the Universe is flat ($\kappa = 0$) and filled with nonrelativistic 
matter, then 
the scale factor evolves as $a\propto t^{2/3}$ (matter domination). 
If the Universe is filled with ultrarelativistic matter, 
then scale factor increases 
as $a\propto t^{1/2}$ (radiation domination).
There is also dark energy in our Universe, whose density does not
change in time (or almost does not change in time), $\rho_\Lambda
= {\rm const}$. The Universe was radiation--dominated at early epoch,
then there was long matter--dominated epoch, and recently dark energy
has started to dominate the cosmological expansion.

At the epochs we consider, the Friedmann equation is specified to
\begin{eqnarray}
H^2 \equiv\left(\frac{\da}{a}\right)^2&=&\frac{8\pi G}{3}
(\rho_{\rm rad}+\rho_{\rm M}+\rho_\Lambda) - \frac{\kappa}{a^2}=\\
&=&\frac{8\pi G}{3}\rho_{\rm c}\left[\Omega_{\rm rad}
\left(\frac{a_0}{a}\right)^4+
\Omega_{\rm M}\left(\frac{a_0}{a}\right)^3
+\Omega_\Lambda + \Omega_\kappa \left(\frac{a_0}{a}\right)^2\right],\nonumber
\label{may29-10}
\end{eqnarray}
where the index ``0'' refers to the current epoch; $\rho_{\rm rad},\rho_{\rm M},\rho_\Lambda$ are energy densities of radiation, nonrelativistic
matter 
(i. e., baryons and dark matter) and dark energy, respectively,
 $\rho_{\rm c}=(8\pi G)/(3H_0^2)$, and $\Omega_{{\rm rad,M},\Lambda}=
\rho_{{\rm rad},M,\Lambda}/\rho_c$ are the present 
energy densities in units of the critical density, while $\Omega_\kappa$
is the relative present contribution of the spatial curvature to the current Universe expansion rate. $\Omega$'s are relative contributions in the sense that $\Omega_{\rm rad}+\Omega_{\rm M}+\Omega_\Lambda+\Omega_\kappa=1$.

The coordinate $t$ is the physical time, it is the proper time of cosmological observers. Sometimes another time coordinate is used, so-called conformal time
\be
\eta(t)=\int\limits^t\frac{dt'}{a(t')}.
\ee
In coordinates $\eta,x_1,x_2,x_3$ the metric (\ref{friedman}) for $\kappa=0$
reads:
\be
ds^2=a(\eta)^2(d\eta^2-d\vecc{x}^2),
\label{may29-5}
\ee
i. e., the metric is conformally flat. Since the light propagation satisfies $ds^2=0$, in coordinates $\eta,x_1,x_2,x_3$ the light propagates as in the Minkowski space--time. 
%\subsection{Initial conditions and evolution of perturbations}
\subsection{The angular spectrum of the CMB temperature anisotropy}
\subsubsection{Sachs--Wolfe, Doppler and integrated Sachs--Wolfe 
effects\label{SW-D-ISW}}
As we have already pointed out, the major source of the CMB anisotropy are the
density perturbations and gravitational potentials associated with 
them\footnote{We do not consider the effects due to structures in 
the relatively
late Universe, such as Sunyaev--Zel'dovich effect and gravitational lensing.
Yet another possible source is gravity waves, see below.}. In cosmological
literature these are called scalar perturbations, since the relevant
quantities are scalars under spatial rotations. In a certain gauge 
(conformal Newtonian gauge), metric with scalar perturbations is written
as follows:
\be
ds^2 = (1+2\Phi)dt^2 - a^2(t)(1+2\Psi)d\vecc{x}^2,
\ee
where $\Phi(\vecc{x},t)$ and $\Psi(\vecc{x},t)$ 
are two gravitational potentials. In the Newtonian 
limit, $\Phi$ coincides with the Newtonian potential.

In the case of scalar perturbations, the observed
temperature fluctuation of the CMB photons coming to us
along the direction $\vecc{n}$ from the point $\vecc{x}$ where they have 
last scattered is given by the general formula\footnote{This 
formula does not account for
aforementioned smearing effects, so it is valid, strictly speaking, 
only at relatively large
angular scales corresponding to $l \lesssim 1500$. However, smearing
leaves almost unaffected some properties of the CMB angular spectrum,
such as positions of the peaks. We use this formula for qualitative analysis
anyway.}:
\be
\frac{\delta T}{T} = \frac{1}{4} \frac{\delta \rho_\gamma}{\rho_\gamma}(\vecc{x})
+ \Phi (\vecc{x}) + \vecc{n} \cdot \vecc{v}_\gamma (\vecc{x}) +
\int\limits_{t_r}^{t_0}~dt ~(\dot{\Phi} - \dot{\Psi}),
\label{may29-1}
\ee
where $\delta \rho_\gamma (\vecc{x})$ is the density perturbation of photons
at last scattering, $\vecc{v}_\gamma (\vecc{x})$ is the velocity of the
baryon--photon medium at that time, and the integral is evaluated along the
photon world line from last scattering to detection. All terms here have
transparent interpretation. The first term is the perturbation of the 
temperature of the photon gas at last scattering: in general $\rho_\gamma \propto T_\gamma^4$ and we have 
$1/4\cdot\delta \rho_\gamma(\vecc{x})/\rho_\gamma 
= \delta T_\gamma(\vecc{x})/T_\gamma$. The second term reflects the
fact that photons get redshifted (blueshifted) when they escape from
potential wells 
(humps). The 
first and second contributions together are called the Sachs--Wolfe effect.
The third term is self-evident: this is
the Doppler effect. Finally,
the integral in (\ref{may29-1}) represents the integrated Sachs--Wolfe effect:
as photons propagate through the Universe, they get red- or blueshifted
due to time-dependent gravitational field. Namely, if a photon meets a 
potential well, it falls into it and gains energy. Suppose now that
by the time the photon starts escaping from the well, the well has become
shallower. Then the photon looses less energy when climbing the well,
and the net effect is the blueshift of the photon. Since photons are
relativistic, this effect is caused by both temporal
($\delta g_{00} \propto \Phi$) and spatial ($\delta g_{ij} \propto
\Psi \delta_{ij}$) components of metric perturbations.

For scalar perturbations, both Doppler and integrated Sachs--Wolfe 
effects are numerically rather small. We simplify our discussion and
consider the Sachs--Wolfe effect alone (the first two terms in 
(\ref{may29-1}). The main contributions here come from perturbations in the
baryon--photon medium and dark matter. It is important that to a good
approximation, baryons and photons make a single fluid before recombination;
they are tightly coupled due to intense scattering of photons off
free electrons and intense Coulomb interaction between electrons and 
protons\footnote{The tight coupling approximation is {\it not} valid
for $l \gtrsim 1500$ because of the Silk damping and other effects.
Our discussion, however, remains valid at qualitative level even
for large $l$.}.

Let us consider the effects due to baryon--photon medium and dark
matter in turn.

\subsubsection{Oscillating part of the angular
spectrum: baryon--photon medium \label{oscsect}}
The oscillating part of the CMB temperature angular spectrum comes from
sound waves in the baryon--photon medium before recombination.
So, we have to understand the behavior of these waves in the expanding
Universe.
Instead of writing and then solving the equations for sound waves, let us consider a toy example, the case of inhomogeneities in massless scalar field. 
In fact,  example is useful for demonstrating the
 main notions and ideas of treatment of 
various kinds of perturbations. The action for the scalar field is:
\be
S=\int d^4x\sqrt{-g}\left[\frac12g^{\mu\nu}\p_\mu\phi\p_\nu\phi\right]=\int d^3xdta^3\left[\frac12\dot{\phi}^2-\frac{1}{2a^2}(\p_i\phi)^2\right].
\ee
The field equation thus reads:
\be
-\frac{d}{dt}(a^3\dot{\phi})+a\p_i\p_i\phi=0,
\ee
i. e.,
\be
\ddot{\phi}+3H\dot{\phi}-\frac{1}{a^2}\p_i^2\phi=0. \label{scalarfield}
\label{may29-3}
\ee
Here, $H$ denotes the Hubble parameter: $H\equiv\da/a$. This equation is linear in $\phi$ and homogeneous in space, so it is natural to represent $\phi$ in terms of the Fourier expansion:
\be
\phi(\vecc{x})=\int e^{i\vecc{k}\vecc{x}}\phi(\vecc{k})d^3k.
\ee
Clearly, the value of $\vecc{k}$ 
for a given Fourier mode is a constant. However, $k$ is not a physical wavenumber (physical momentum), since $x$ is not a physical distance. $k$ is called conformal momentum, while physical momentum equals $q\equiv 2\pi/\lambda=2\pi/(a(t)\Delta x)=k/a(t)$. $\lambda$ here is the physical wavelength of perturbation, and it grows together with the expansion of the Universe. Accordingly, as the Universe expands, the physical momentum of a given mode decreases (gets redshifted), $q(t)\propto 1/a(t)$. For a mode of given conformal momentum $k$, Eq.(\ref{scalarfield}) gives:
\be
\ddot{\phi}+3H\dot{\phi}+\frac{k^2}{a^2}\phi=0.
\ee
This equation has two time--dependent
parameters of the same dimension: $k/a$ and $H$. One can consider two limiting cases: $k/a\ll H$ and $k/a\gg H$. In the 
cosmological models
with conventional equation of state of the dominant component (e. g., matter--dominated or radiation--dominated Universe), one has
$H\approx 1/L$, where $L$ is the size of cosmological horizon. So, the regime $k/a \ll H$ is the regime when the physical wavelength is greater than the horizon size (super-horizon mode), while for $k/a \gg H$ the physical wavelength is much smaller than the horizon size (sub-horizon mode). The time when the wavelength of the mode coincides with the horizon size is called horizon crossing. This
time is denoted by the
symbol $\times$. It is straightforward to see that at both radiation--dominated and matter--dominated epochs, the ratio $k/(aH)$ grows. This means that every mode was at some early time super-horizon, and later it becomes sub-horizon.

%\subsubsection{Evolution of super-horizon mode in radiation--dominated regime}
For a super-horizon mode, we can neglect the term $\phi\cdot k^2/a^2$ in Eq.(\ref{scalarfield}) in comparison with other terms. Then the field equation, 
e.g., in
the radiation--dominated Universe ($a\propto t^{1/2},H=1/2t$), becomes
\be
\ddot{\phi}+\frac{3}{2t}\dot{\phi}=0.
\ee
The solution of this equation is
\be
\phi(t)=A+\frac{B}{\sqrt{t}}.
\ee
This behavior is generic for all cosmological perturbations: there is 
a constant mode ($A$ in our case) and a mode that decays in time.
If we extrapolate the decaying mode $B/\sqrt{t}$ back in time, we 
get very strong (infinite in the limit $t\rightarrow 0$) perturbation. 
For density perturbations this means that this mode corresponds to
strongly inhomogeneous early Universe. 
Therefore, this mode is usually assumed to be absent for actual 
cosmological perturbations. 
Hence, for given $\vecc{k}$, the solution is determined by
a single parameter, the initial amplitude $A$ of the mode
$\phi_{\vecc{k}}$. 

After entering the sub-horizon regime, the modes begin to oscillate.  The equation in the sub-horizon regime is:
\be
\ddot{\phi}+\frac{3}{2t}\phi+\frac{k^2}{t}\phi=0,
\ee
and its general solution in the WKB-approximation reads:
\be
\phi(t)=\frac{A'}{a(t)}\cos\left(\int\limits_0^t\frac{k}{a(t')}dt'+\psi_0\right),
\label{may29-2}
\ee
with the two constants being the amplitude $A'$ and the phase $\psi_0$. The amplitude $A'$ of these oscillations is determined by the amplitude of the super-horizon initial perturbation, while {\it
the phase $\psi_0$ of these oscillations 
is uniquely determined by the condition of the absence of the decaying mode}, $B=0$. Imposing these conditions yields
\be
\phi(t)=D\frac{a_{\times}}{a(t)}\sin\left(\int\limits_0^t\frac{k}{a(t')}dt'\right).
\label{may31-7}
\ee
The decreasing amplitude of oscillations $\phi(t) \propto 1/a(t)$ and the particular phase $\psi_0 = -\pi/2$ in (\ref{may29-2}) are
peculiar properties of the wave equation (\ref{may29-3}), as well as
the radiation--dominated cosmological expansion. However,
{\it the fact that the phase of oscillations
is uniquely determined by the requirement
of the absence of the super-horizon decaying mode is generic}.

The perturbations in the baryon--photon medium before recombination 
--- sound waves --- behave in a rather 
similar way.  Their evolution is as follows:
\be 
\frac{\de\rho_\gamma}{\rho_\gamma}=\left\{\ba{cc} {\rm const}, & {\rm outside \quad horizon}, \\
{\rm const}\cdot \cos\left(\int\limits_0^tv_s\frac{k}{a(t')}dt'\right), & {\rm inside \quad horizon}, \ea \right.
\ee
where $v_s\equiv \sqrt{d p/d\rho}$ is the sound speed. The  baryon--photon medium before recombination is almost relativistic\footnote{This does not contradict the statement that the Universe is in matter--dominated regime at recombination. The dominating component at this stage is dark matter.} since $\rho_{\rm B}<\rho_\gamma$. Therefore, $v_s\approx 1/\sqrt{3}$. At the time of last scattering $t_r$ we have:
\be
\frac{\de\rho_\gamma}{\rho_\gamma}={\rm const} \cdot \cos\left(\int\limits_0^{t_r}v_s\frac{k}{a(t')}dt'\right)={\rm const}\cdot \cos kr_s,
\label{may30-5}
\ee
where 
\be
r_s=\int\limits_0^tv_s\frac{dt'}{a(t')}
\ee
 is the comoving size of the sound horizon, while the physical size equals $a(t_r)r_s$. So, we see that the value 
of the
density perturbation at recombination oscillates as a function of wavelength. The period of this oscillation is determined by $r_s$.
These oscillations are shown in Fig. \ref{anisotropyspectrum}.
% We observe these oscillations as oscillations of the CMB angular spectrum as function of the multipole number $l$.
 Indeed, one can show that gravitational potentials
induced by the perturbations in the baryon--photon medium are negligible
at recombination, while, as we mentioned above, the Doppler and 
integrated Sachs--Wolfe effects are also rather small. 
Hence, the main effect of the baryon--photon
perturbations on CMB comes from the oscillatory first term in 
(\ref{may29-1}).  The absolute values of perturbations\footnote{According to
(\ref{may29-6}), the relevant quantity is the temperature fluctuation squared.
This is the reason for considering absolute values of density
perturbations.} 
in the baryon--photon plasma are maximal at conformal 
momenta $k_n$ obeying $k_nr_s=\pi n$ for integer $n$. Hence, their 
coordinate half wavelengths (distances between maxima of the absolute value)
are equal to $\Delta x_n = \pi/k_n$. Now, according to
(\ref{may29-5}), photons propagate in conformal coordinates
$(\eta, \vecc{x})$ precisely like in the Minkowski space--time. Hence,
an interval of coordinate length $\Delta x$ at the time of last
scattering is seen today at an angle $\Delta x/\eta_0$, where $\eta_0$
is the conformal time interval between the present and last scattering epochs,
i. e., the comoving distance to the sphere of last 
scattering\footnote{This is true for intervals normal to the line
of sight. One can show that the sound waves in the baryon--photon component,
 whose wave vectors are normal to the line of sight, indeed give dominant
contribution to the oscillations in the CMB temperature angular
spectrum. Contributions to the oscillating part of the angular spectrum
coming from other 
waves get smeared out upon integration over the directions of
the wave vectors.}.
So, the strongest CMB temperature fluctuations have angular scales
$\Delta \theta_n \simeq r_s/(n \eta_0)$.
We recall the approximate correspondence 
$\Delta \theta \longleftrightarrow \pi/l$ and conclude that
the peaks in the CMB angular spectrum are expected to be at the positions
\be
l_n \simeq \pi n\frac{\eta_0}{r_s}. \label{lmax}
\ee
 These peaks are indeed seen in
% Fig.\ref{anisotroptopdef} and
 Fig.\ref{anisotropyspectrum}. 

Let us pause here to record the approximate correspondence between the
conformal momenta of perturbations and the multipole numbers of the
CMB anisotropy they generate:
\be
l \longleftrightarrow k\eta_0.
\label{may30-2}
\ee
This correspondence follows from the argument just presented and
will be used later on.

The very existence of the peaks in the CMB angular spectrum strongly suggests
that {\it the 
hot Big Bang stage of the cosmological expansion was preceded by
another epoch with entirely different properties.} That was the epoch
when the cosmological perturbations were generated.
Indeed, we have seen
that the definite phases of sound waves which are behind the peaks are
determined by the properties of the scalar perturbations in the early,
super-horizon regime. So, the modes we observe were indeed super-horizon,
their wavelengths  
exceeded considerably the size of the cosmological horizon. In the hot Big
Bang theory, there is no causal connection between horizon-size regions.
Hence, within this theory it is impossible to invent a mechanism 
which would generate these perturbations in a causal manner. We conclude that
the generation of the cosmological perturbations must have
occurred at some epoch
preceding the hot Big Bang stage. Furthermore, the cosmological evolution at
that epoch was such that the combination $a(t)H(t)$ was small early on, 
and then it increased enormously towards the beginning of the
hot Big Bang stage. Only in that case the physical momentum $k/a(t)$ could
exceed the Hubble parameter $H$ at early times, the modes could be sub-horizon
``in the beginning'', and the perturbations could be generated by some causal
mechanism. This is precisely what happens in the inflationary scenario. 

\begin{figure}[h!]
\begin{center}
\includegraphics[width=0.6\linewidth]{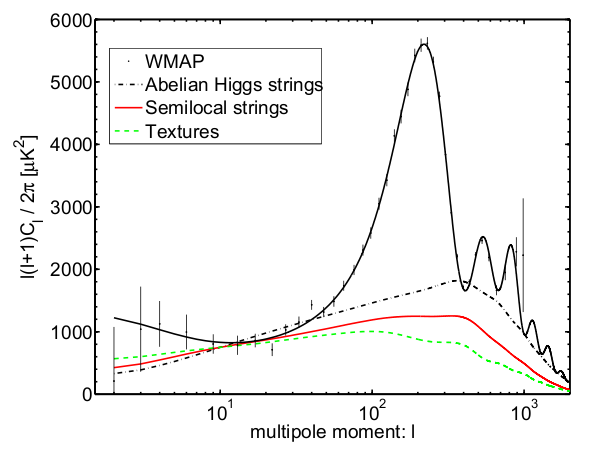}
\caption{~The real CMB angular spectrum and
inflationary prediction compared with the spectra obtained in models
with casual mechanisms of the generation of density perturbations
\cite{strings}. In all casually connected models (all
models above except inflation), there are no oscillations.}
\label{anisotroptopdef}
\end{center}
\end{figure}

An alternative would be the generation of the density perturbations at
the hot Big Bang stage in a causal way. This can only happen when a mode
of perturbations is in
the sub-horizon regime. 
In that case, there is no reason for
very definite phases of the sound waves in the baryon--photon component,
and hence for oscillations in the CMB temperature angular spectrum. 
For example, Fig.\ref{anisotroptopdef}  shows the
angular spectrum in a model where density perturbations 
are generated by accretion on cosmic strings, and the
spectra created by other casual mechanisms at the hot
Big Bang epoch. All these spectra do not  
exhibit the oscillatory behavior. So, the generation of density perturbations
at the hot Big Bang stage is ruled out by the CMB observations. 
%Today there are no ideas of how casual-connected mechanism can produce such a spectrum.
% So, we can conclude that the inhomogeneities were formed before the stage of hot Big Bang.
% In two words, the inhomogeneities were formed before the stage of hot Universe (??????). 

The positions and strengths of 
the peaks are sensitive to
various cosmological parameters, so from observations one 
can determine these parameters or at least their combinations. 
In particular, {\it the peak positions}  depend on:
\begin{enumerate}
\item The ratio of baryons to photons. 
\item The rate of the cosmological expansion, thus the density of dark energy and other forms of matter.
\item The curvature of space. 
\end{enumerate}
 The size of the sound horizon entering (\ref{lmax}) depends on the sound speed in 
the baryon--photon component. It is given by
\be
v_s^2=\frac{d p}{d \rho}=\frac13\frac{d\rho_\gamma/\rho_\gamma}{\frac{d\rho_{\rm B}}{\rho_{\rm B}}\frac{\rho_{\rm B}}{\rho_\gamma}+\frac{d\rho_\gamma}{\rho_\gamma}}=\frac13\frac{4dT/T}{3\frac{\rho_{\rm B}}{\rho_\gamma}\frac{dT}{T}+4\frac{dT}{T}}=\frac13\frac{1}{1+\frac34\frac{\rho_{\rm B}}{\rho_\gamma}}.
\ee
The dependence of $v_s^2$ and hence $r_s$ on $\rho_{\rm B}$ is rather weak,
 since $\rho_{\rm B}/\rho_\gamma \ll 1$ up to recombination. 
More sensitive to the baryon fraction are the heights of the peaks,
see below. For given baryon abundance, precisely determined basically
from the peak positions, the sound horizon serves as a standard ruler at
recombination. The angle at which it is seen today depends on the 
expansion history after recombination and spatial curvature.

The dependence on the expansion history
of the Universe comes about in the following way. 
We know precisely the factor $a(t_0)/a(t_r)$ determining how much
the Universe has 
expanded from the time of last scattering 
(since   $a(t_0)/a(t_r) = T_r/T_0$ and 
we know the present CMB temperature $T_0$ 
and the temperature of last scattering $T_r$). The Universe could expand fast, and then the last scattering surface would be close to us, or it could expand slowly, and the last scattering surface would be further away, 
so the angular scale at the 
peaks would be smaller. Notably, 
the expansion rate depends strongly on the value of the cosmological
constant $\Lambda$ (or, generally, on dark energy density); we discuss this
point shortly in some detail.

\begin{figure}[h!]
\begin{center}
\includegraphics[width=0.95\linewidth]{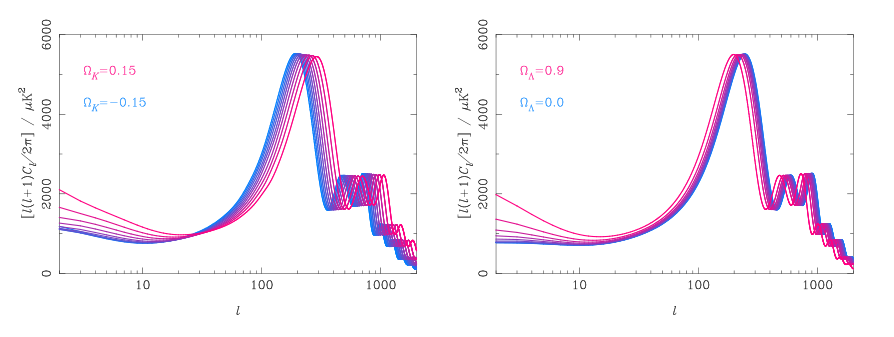}
\caption{~The effect of curvature (left panel) and of $\Lambda$ (right panel) on the CMB temperature angular spectrum \cite{Challinor}.}
\label{curvlambdaeffect}
\end{center}
\end{figure}

The dependence on curvature of space comes from the fact that the same interval viewed from the same distance is seen at different angles depending on curvature. For example, in the space of positive curvature (3-sphere) the angle is larger than the one in the Euclidean space, and the angle in the flat space is larger that the angle in the space of negative curvature (Lobachevsky geometry). So, both the curvature and the cosmological constant lead to similar shift of the peaks in the CMB angular spectrum, see Fig. \ref{curvlambdaeffect}. Therefore, there is a degeneracy in the parameters of $\Omega_\Lambda$ and $\Omega_\kappa$,
% which is illustrated in Fig.~\ref{curvlambda}. Figure~\ref{curvlambda}
see   \cite{WMAPKom}. Actually, the degeneracy is lifted by making use of the results of cosmological observations unrelated to CMB. We give an example of the latter in Section~\ref{sec:BAO}.

Let us illustrate the dependence of peak positions on the dark energy density by considering the spatially
flat Universe. The Friedmann equation
(\ref{may29-10}) in the flat Universe can be written as follows:
\be
\left(\frac{\da}{a}\right)^2
=\frac{8\pi G}{3}\rho_0\left(\Omega_{\rm rad}(1+z)^4+\Omega_{\rm M}(1+z)^3+\Omega_\Lambda \right), \label{fried}
\ee
where 
$z=a_0/a(t)-1$ is redshift. As we have already noted,
 the light propagates according to the
equation $ds^2=0$, i. e., $dt^2=a^2dx^2$, so 
$\eta=\int(1/a)dt$ is the 
comoving distance traveled 
by light. To calculate the comoving distance which light has 
traveled since recombination, one uses Eq.~(\ref{fried}) and writes
% Solving the equation (\ref{fried}):
\be
dz=-\frac{a_0}{a^2}da=-\frac{a_0}{a}\frac{\da}{a}dt=-a_0 H\frac{dt}{a}. 
\ee
Therefore,
\be
\eta_{0}=\int\limits_{t_r}^{t_0}\frac{dt}{a}=\int\limits_0^{z_r}\frac{dz}{a_0H(z)}=\frac{1}{a_0H_0}\int\limits_0^{z_r}\frac{dz}{\sqrt{\Omega_{\rm rad}(1+z)^4+\Omega_{\rm M}(1+z)^3+\Omega_\Lambda}}.
%\eta=\int_{t_r}^{t_0}\frac{dt}{a}=\int_0^{z_r}\frac{dz}{a_0H(z)}=\frac{1}{a_0H_0}\int_0^{z_r}\frac{dz}{\sqrt{\Omega_{\rm rad}\left(\frac{a_0}{a}\right)^4+\Omega_{\rm M}\left(\frac{a_0}{a}\right)^3+\Omega_\kappa\left(\frac{a_0}{a}\right)^2+\Omega_\Lambda}}
\ee
%The inhomogeneity of physical size $d$ at last scattering is seen at the angle of $\theta=\frac{d}{a_r \eta_{0r}}$ 
The effect on $\eta_0$ due to $\Omega_\Lambda$ is seen from this formula. Note that by definition $\Omega_{\rm rad}+\Omega_{\rm M}+\Omega_\Lambda=1$ in the spatially
flat Universe, 
so the larger $\Omega_\Lambda$ is, 
the larger is the integrand at $z>0$. Since the positions of the
peaks are approximately given by Eq.(\ref{lmax}), these positions 
are shifted to the left at larger $\Omega_{\Lambda}$. 
\subsubsection{Nonoscillating part of the spectrum: the effect of dark matter}
As we saw above, 
matter with finite sound speed  produces oscillating angular
spectrum of the CMB temperature anisotropy. 
But there exists another form of matter, which has  
zero (or nearly zero)
sound speed. This is dark matter. Due to the fact that $p=0$ for dark matter,
and hence $v_s^2=0$, 
there are no waves in this medium, and thus by itself it does not 
produce oscillations in the CMB angular spectrum. 
Nevertheless, dark matter density perturbations
affect the CMB angular spectrum quite strongly, via the
% Now we consider the influence of dark matter on the spectra of CMB inhomogeneities.
% Dark matter affects the baryon--photon medium by creating 
gravitational potentials they create. These gravitational potentials
induce the CMB temperature anisotropy both directly, through the second term
in (\ref{may29-1}), and through the induced perturbations in the 
baryon--photon medium, which falls into the potential wells.
%, and we see more dense regions. 

In the first place, we have to understand the properties of the
gravitational potential $\Phi_{\rm DM}$ generated by dark matter
perturbation.
Let us consider a  perturbation whose wavelength is much smaller than the
horizon size at recombination. Then the Newtonian approximation is valid,
and the gravitational potential is related to the mass density perturbation
by the Poisson equation. We take into account the fact that the physical
Laplacian written in terms of comoving coordinates is 
$\Delta = a^{-2} \partial^2/\partial \vecc{x}^2$ and write 
the Poisson equation in the expanding Universe in the following way
(we sometimes omit subscript DM in $\Phi_{\rm DM}$ in this Section):
\be
-\frac{k^2}{a^2 (t)}\Phi=4\pi G \de\rho_{\rm DM}.
\ee
In matter--dominated regime $\de\rho_{\rm DM}/\rho_{\rm DM}\propto a(t)$ for sub-horizon 
modes, while $\rho_{\rm DM} \propto 1/a^3$. Therefore, 
$\de\rho_{\rm DM}\propto 1/a^2$ and $\Phi$ {\it is
independent of time}. 
Using the assumption of the
flat spectrum of primordial perturbations, for which $\de\rho_{\rm DM}/\rho_{\rm DM}$ 
depends on $k$ weakly (see below), we obtain: 
\be
\Phi_{\rm DM} (k) \propto \frac{1}{k^2}.
\label{may30-1}
\ee
In fact, this relation is valid modulo logarithm, whose origin is the
evolution of the dark matter perturbations at the radiation--dominated
epoch; we are not going to elaborate on this point.

To evaluate the effect of dark matter on CMB, we have to obtain
the photon density perturbation $\delta \rho_\gamma/\rho_\gamma$,
which is induced by
the gravitational potential $\Phi_{\rm DM}$.
In this way we are going to find the first term in   (\ref{may29-1});
as we have already pointed out, the third and fourth terms are rather small,
and we neglect them for the sake of argument.
To this end, let 
us consider the baryon--photon 
medium in static background gravitational potential. Baryon-photon plasma 
is then in the state
of hydrostatic equilibrium.
In Newtonian mechanics, the equation of hydrostatic equilibrium is:
\be
\vecc{\nabla}p=-\rho\vecc{\nabla}\Phi.
\ee
In the case of relativistic medium, when 
$p$ is of order $\rho$ (but  $\Phi$ is small,  $\Phi\ll 1$), 
this equation is generalized to
\be
\vecc{\nabla} p =-(p+\rho)\vecc{\nabla}\Phi,
\ee
or, in the Fourier space:
\be
\de p=-(p+\rho)\Phi.
\ee
Here $p$ and $\rho$ are pressure and energy density of the baryon--photon
medium before recombination.
Using the equation of state $p_\gamma=\rho_\gamma/3$, $p_{\rm B}=0$ we obtain:
\be
\frac13\de\rho_\gamma=-\left(\frac43\rho_\gamma+\rho_{\rm B}\right)\Phi, \label{drhogamma}  
\ee
and, finally,
%Now, $\rho_\gamma\propto T_\gamma^{4}$ and hence
%\be
%\frac{\de T_\gamma}{T_\gamma}=\frac14\frac{\de\rho_\gamma}{\rho_\gamma}  \label{stefboltz}
%\ee
%This gives
\be
\frac14\frac{\de\rho_\gamma}{\rho_\gamma}=-\left(1+\frac34\frac{\rho_{\rm B}}{\rho_\gamma}\right)\Phi. \label{dtt}
\ee

Let us pause here to point out that
there is a subtlety in this result. Namely, suppose for the moment that 
baryon density is negligible, and consider static flat background
space--time. Let there be dark matter which does not interact with photons, but creates the gravitational potential. Then in thermodynamic equilibrium the temperature of photons should be the same everywhere in space.
This is in apparent contradiction with the result (\ref{dtt}), since
the left hand side is precisely the perturbation of the photon temperature:
$\delta \rho_{\gamma}/\rho_\gamma = 4\delta T_\gamma/T_\gamma$
in view of the general property $\rho_\gamma \propto T_\gamma^4$.
The reader is invited to reconcile thermal equilibrium with the result 
(\ref{dtt}).

Coming back to CMB, we find from Eq.~(\ref{dtt}) that the Sachs--Wolfe
effect on the observed CMB temperature --- the first two terms in
(\ref{may29-1}) --- is
\be
\frac{\de T}{T}=-\frac34\frac{\rho_{\rm B}}{\rho_\gamma}\Phi_{\rm DM}.
\ee
Notably, the net effect of dark matter is proportional to
$\rho_{\rm B}$:
if there were no baryons, the gravitational potential would not affect 
CMB. 

According to (\ref{may30-2})  and (\ref{may30-1}), the dark matter contribution
to the CMB angular spectrum is a smooth function of $l$ that decays
as $1/l^2$. To large extent this contribution determines
the overall decline of the spectrum, with oscillations superimposed on
it, which is clearly seen in Fig.~\ref{anisotropyspectrum} (our Newtonian
analysis is not valid for perturbations that remain super-horizon at 
recombination, so our discussion applies to the region $l \gtrsim 100$
only).

\begin{figure}[h!]
\begin{center}
\includegraphics[width=1.0\linewidth]{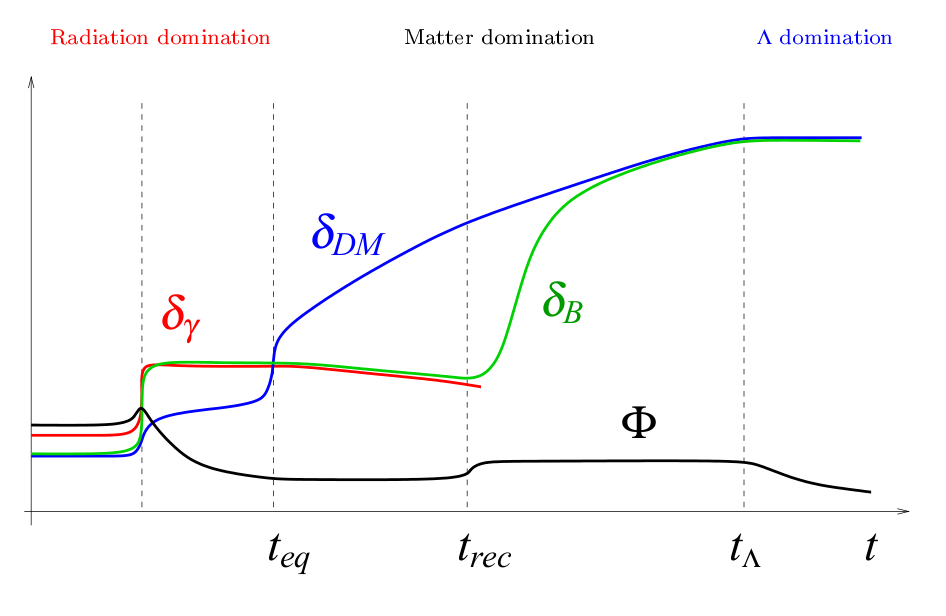}
\caption{~Schematic plot of
the evolution of density pertirbations in different components.
Here, $\de=\de\rho/\rho$, and $\Phi$ is the gravitational potential. The 
left dashed vertical line is the time of horizon crossing of a mode
considered.}
\label{modegrowth}
\end{center}
\end{figure}

The effect of dark matter on CMB is so strong that the existence
of dark matter could be deduced from CMB observations even in the absence
of astrophysical evidence. As a digression, we point out that
the presence of dark matter is crucial also
for structure formation in the Universe. Indeed, at matter--dominated epoch,
sub-horizon
density perturbations in nonrelativistic matter (e. g., in baryons) grow as $\de\rho/\rho \propto a$. At recombination, inhomogeneities in the baryon component
were of order $10^{-5}$, the Universe since then
has the expansion factor $a(t_0)/a(t_r)=z_r+1\approx 1000$, so the 
perturbations at present would be of order $10^{-2}$. 
They would not have left the linear regime of evolution, and thus would not form galaxies, stars, planets, men, etc. What saves the day is the property
that the perturbations in dark matter grow well before recombination, and soon after recombination baryons catch up with them, 
as shown in Fig.~\ref{modegrowth}.

\subsubsection{Interference}
The sum of
the contributions of the baryon--photon and dark 
matter perturbations to the CMB temperature fluctuation has the following
form:
\be
\frac{\de T}{T} =A\cdot 
\cos(kr_s)-\frac34\frac{\rho_{\rm B}}{\rho_\gamma}\Phi_{\rm DM}.
\label{may30-3}
\ee
The first term comes from the baryon--photon acoustic oscillations, and the second one from dark matter perturbations. 
According to (\ref{may29-6}), the observables are 
quadratic in $\delta T/T$, so an important issue is 
whether these two terms add up coherently. If they are coherent, we have:
\be
\left(\frac{\de T}{T}\right)^2=A^2 \cos^2(kr_s)-\frac{3}{2}\frac{\rho_{\rm B}}{\rho_\gamma}\cos(kr_s)A\Phi_{\rm DM}+\left(\frac34\frac{\rho_{\rm B}}{\rho_\gamma}\Phi_{\rm DM} \right)^2.
\label{jun1-2}
\ee
Suppose now that the relative sign of $A$ and $\Phi_{\rm DM}$ is positive.
Then
the peaks corresponding to $\cos kr_s =-1$ (odd $n$ in (\ref{lmax}))
are enhanced, and the peaks corresponding to
$\cos kr_s =1$ (even $n$) are suppressed due to the second,
interference term. 
Since $\Phi\propto 1/k^2$, the effect is more pronounced for the first peaks. 
And this is exactly what is observed. 

Let us recall that the density perturbations are random fields.
Hence, the interference is possible only if the baryon--photon
perturbations (the coefficient $A$ in (\ref{may30-3}))
and dark matter perturbations ($\Phi_{\rm DM}$ in  (\ref{may30-3}))
are proportional to one and the same random field, call it $\R(\vecc{k})$.
This means that {\it the baryon--photon and dark matter perturbations
are of one and the same origin}, with $\R$ being the common
amplitude of primordial scalar perturbation.
In fact, CMB data are consistent with the property that primordial
scalar perturbations are in {\it the adiabatic mode}.
The 
definition of the
adiabatic mode is that the particle content is one and the same throughout
the Universe. In other words, adiabatic mode would appear if one contracts or expands some regions of the Universe without changing the 
chemical composition
of matter in these regions. The invariant and time independent characteristic
of the baryon abundance is the ratio $n_{\rm B}/s$ of the baryon number density
to the entropy density. Hence, in the adiabatic mode $n_{\rm B}/s$
is constant in time {\it and space}. Likewise, the ratio of
the number density of dark matter particles to the entropy density
$n_{\rm DM}/s$ is a universal constant. Since $s \propto T^3$ and 
$\rho_\gamma \propto T^4$,
for the adiabatic mode in super-horizon regime we have\footnote{The 
relations (\ref{may31-2}) are valid in the conformal
Newtonian gauge. A
convenient gauge-invariant
definition of $\R$ is the spatial curvature of comoving 
hypersurfaces.}:
\be
\frac{\de\rho_{\rm DM}}{\rho_{\rm DM}}=\frac{3}{4}\frac{\de\rho_\gamma}{\rho_\gamma}=\frac{\de\rho_{\rm B}}{\rho_{\rm B}}=\frac34\frac{\de\rho_\nu}{\rho_\nu} =\R(\vecc{k}).
\label{may31-2}
\ee
After the horizon entry, the perturbations in the baryon--photon and
dark matter components start to evolve. This evolution is linear
(perturbations are small) up until recombination, so we have at
recombination
\be
\Phi_{\rm DM} (\vecc{k})=T_{\rm DM}(k)\R(\vecc{k}), \label{tdm} \\
A(\vecc{k})=T_{B\gamma}(k)\R(\vecc{k}). \label{tbg}
\ee
The functions $T_{\rm DM}$ and $T_{\rm B\gamma}$ are called transfer functions. 
They describe how the perturbations 
in different media evolve. Yet $\Phi_{\rm DM}$ and $A$ at recombination
are proportional to
one and the same random field $\R$, and they add up coherently.

\begin{figure}[h!]
\begin{center}
\includegraphics[width=1.0\linewidth]{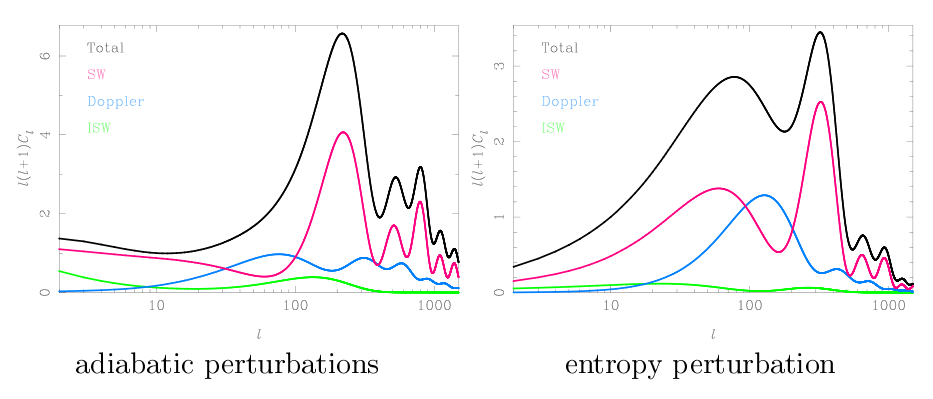}
\caption{~The effects of adiabatic and entropy perturbations on the CMB  angular spectrum  \cite{Challinor}.}
\label{adiabentr}
\end{center}
\end{figure}

In principle, besides the adiabatic mode, there could
exist  so-called CDM entropy mode and baryon entropy mode.
In the CDM entropy mode, there are no primordial
perturbations in the baryon--photon medium, 
perturbations exist only in dark matter. 
Such a situation persists until the horizon crossing, due to causality.
The inhomogeneities in the baryon--photon medium 
are generated at the time of horizon crossing, 
and afterwards they oscillate as follows: 
\be
\frac{\de \rho_{\gamma}}{\rho_\gamma}
\propto \sin\left(\int\limits_{t_{\times}}^tv_s\frac{k}{a(t')}dt'\right).
\ee
We  emphasize that the phase of oscillations is different for
adiabatic perturbations (cosine rather than sine, see
(\ref{may30-5})). So, if primordial perturbations were in the
CDM entropy mode, the peaks in the CMB angular 
spectrum would be shifted, see Fig.~\ref{adiabentr}. 
Exactly the same result holds for the baryon entropy mode.
 Presently, the admixture of the CDM entropy perturbations is ruled out by 
the CMB observations  at the level of about $10\%$.

Purely adiabatic mode of perturbations is very natural.
If our Universe was in complete thermal equilibrium (including
chemical equilibrium) at some early stage of the hot Big Bang epoch,
the only mode of perturbations is adiabatic. Physical processes
that generated dark matter and baryon asymmetry, if that happened at the
hot Big Bang epoch, were the same everywhere in space, so
the values $n_{\rm DM}/s$ and $n_{\rm B}/s$ are the same everywhere\footnote{The argument 
is, in fact, even more general: the adiabatic mode is the only mode also
in the Universe where
particles of all types are decay products of one and the same field, like
inflaton in the inflationary scenario.}. 
 Reversing the
argument,
if there exists an entropy mode in our Universe, the generation of the
 baryon asymmetry and/or dark matter must have happened before
 the  hot Big Bang epoch. Hence, 
the detection of any of the entropy modes would be a strong signal
for rather unconventional cosmology.

To end up with the interference, we
note that the effect is proportional to $\rho_{\rm B}$ and hence $\Omega_{\rm B}$. 
The same holds for the third, nonoscillating term in (\ref{jun1-2}),
which gives particularly strong contribution in the region of the
first peak.

\begin{figure}[h!]
\begin{center}
\includegraphics[width=0.6\linewidth]{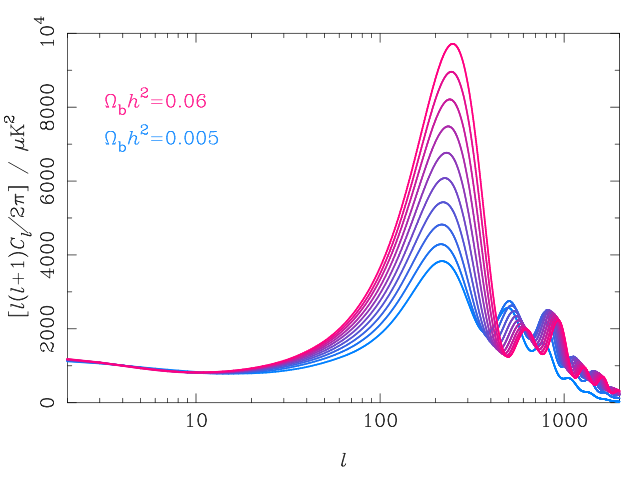}
\caption{~The effect of baryons on the CMB anisotropy spectrum  \cite{Challinor}.}
\label{cmbbaryons}
\end{center}
\end{figure}

In this way the CMB temperature anisotropy is very sensitive
to
$\rho_{\rm B}/\rho_\gamma$ and, therefore, to
the 
baryon asymmetry parameter $\eta_{\rm B}=n_{\rm B}/n_\gamma$ 
and $\Omega_{\rm B}$. 
This is shown in Fig.\ref{cmbbaryons}.
The CMB result for the latter  is:
\be
\Omega_{\rm B}=0.045\pm 0.003.
\ee
Notably, it is consistent with the determination of $\eta_{\rm B}$ from
Big Bang Nucleosynthesis.

\section{Baryon acoustic oscillations}
\label{sec:BAO}

We now digress from the discussion of CMB and introduce one of the
independent methods of measuring combinations of cosmological parameters.
This method has to do with a peculiar property of matter density
distribution in the Universe.

At the time right after  recombination,
the total matter density contrast (dark matter and baryons) is
\be
\frac{\de\rho}{\rho}=\frac{\de\rho_{\rm B}+\de\rho_{\rm DM}}{\rho_{\rm B}+\rho_{\rm DM}}=\frac{\Omega_{\rm B}}{\Omega_{\rm M}}\frac{\de\rho_{\rm B}}{\rho_{\rm B}}+\frac{\Omega_{\rm DM}}{\Omega_{\rm M}}\frac{\de\rho_{\rm DM}}{\rho_{\rm DM}}.
\label{may31-1}
\ee
As we have seen above, right after recombination $\de \rho_{\rm B}$ 
oscillates as a function of
 $k$, while $\de \rho_{\rm DM}$ is a smooth function of $k$. 
After recombination, the matter density distribution is essentially
constant in time (in comoving coordinates), 
because baryons have decoupled from photons, and the
sound speed in the baryon component is essentially zero.
Therefore, the power spectrum of
the total matter density perturbations today has an oscillating in $k$
part. 
This phenomenon is called baryon acoustic oscillations (BAO). 
In the adiabatic mode, the two terms in (\ref{may31-1})
add up coherently, 
so there is an interference between them  in 
$(\de\rho/\rho)^2$. 
The interference is important, since if it did not exist, the oscillations would be the second order effect in the small parameter
$\Omega_{\rm B}/\Omega_{\rm M}$. Due to the interference, the oscillations are of the
first order in $\Omega_{\rm B}/\Omega_{\rm M}$:
\be
\left(\frac{\de\rho}{\rho}\right)^2=\frac{\Omega_{\rm DM}}{\Omega_{\rm M}}B(k)+
\frac{\Omega _{\rm B}}{\Omega_{\rm M}} C(k)\cos(kr_s). \label{BAO}
\ee
This effect has been 
detected in the analysis of the distribution of galaxies
in the Universe at rather small redshifts, $z \simeq 0.2$ and $z \simeq 0.35$,
see \cite{BAO}.

\begin{figure}[h!]
\begin{center}
\includegraphics[width=0.6\linewidth]{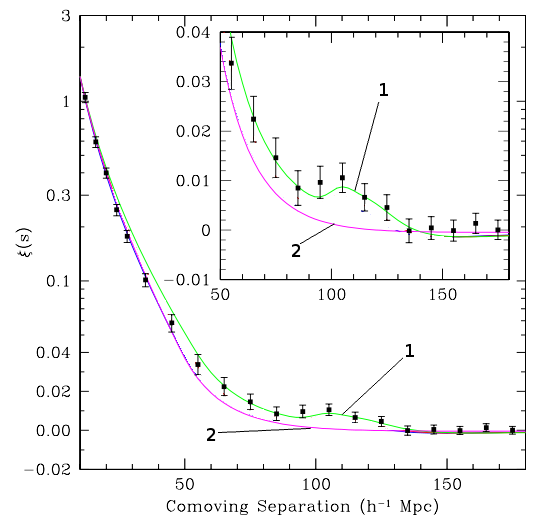}
\caption{~The correlation function of matter  density 
versus distance \cite{BAOPeak}. (1) - The correlation function in 
the model with baryons and $CDM$,  (2) - the correlation function  in pure 
$CDM$-model without baryons.}
\label{corfunc}
\end{center}
\end{figure}

There is a simple interpretation of the effect. In the
adiabatic mode, the overdensities in the
baryon--photon medium and in the dark matter are at the same place before horizon crossing. But before recombination the sound speed in baryon--photon plasma is 
of the order of the speed of light, while the sound speed in dark matter is 
basically zero.
So, the overdensity in baryons generates the outgoing
density wave 
after horizon crossing. 
This wave propagates until  recombination, and then freezes out. 
On the other hand, the overdensity in the dark matter remains in its 
original place.
The current distance from the overdensity in dark matter
to the front of the baryon density 
wave equals\footnote{We remind that the separation at Fig.\ref{corfunc} is given in $h^{-1}$ Mpc, where $h=H/100\quad {\rm km/(s\cdot Mpc)}$, H is a current Hubble parameter. Currently $h\approx 0.7$, so 100 $h^{-1}$ Mpc roughly corresponds to 150 Mpc.} 150 Mpc. Hence, there is an enhanced
correlation between matter perturbations
at this distance scale, which shows up as a feature in the correlation function,
see Fig.~\ref{corfunc}. 
In the Fourier space, this feature produces oscillations \cite{BAO}.

Since the sound speed in the baryon--photon medium before recombination
is well known, these 150~Mpc is a robust prediction, so
BAO provides a standard ruler at low redshifts.
In principle, it can be used to determine the Hubble parameter as
function of $z$ at low $z$ (if the ruler is directed along the line
of sight, its absolute length $l$ is related to its measured extension  in
redshift  by
$\Delta z=H(z) l$) and also its angular diameter distance $D_a(z)$
(if the ruler is normal to the line of sight, the angle at which
it is seen is $\Delta \theta = l/D_a(z)$). The latter again depends on
the expansion history (now at low $z$) and spatial curvature.
In practice, one determines from observations a combination 
$[D_a^2(z)/H(z)]^{1/3}$, which is a rather complicated function of
$\Omega_\Lambda$ and $\Omega_\kappa$. Importantly, this is a different
combination as compared to the one best measured in the CMB observations.
Hence, BAO are used to lift the degeneracy in parameters inherent in the CMB 
data, see   \cite{WMAPKom}.
% as shown in Fig.~\ref{curvlambda}.

\section{Primordial perturbations}
\subsection{Scalar power spectrum and scalar tilt} 
We have seen in Section~\ref{oscsect} that the oscillations in the CMB 
angular spectrum can be naturally explained if we assume that the scalar
perturbations existed 
already at the beginning of the hot Big Bang epoch of 
the cosmological expansion. Also we mentioned that the scalar perturbations 
are observed to be Gaussian, and thus all information about them is contained in the two-point correlation function:
\be
\langle\R(\vecc{k})\R^*(\vecc{k'})\rangle=\frac{P(k)}{(2\pi)^3}\de(\vecc{k}-\vecc{k'}),
\ee
where $\R(k)$ parametrizes the value of the initial perturbation, and $P(k)$ is called power spectrum. In the
isotropic situation, $P(k)$ depends only on the magnitude of vector $k$, but not on its direction. The mean square of fluctuation is
\be
\langle\R^2({\vecc x})\rangle=\langle\int e^{i \vecc{k}\vecc{x}} R(\vecc{k}) d^3k \int e^{-i\vecc{k'}\vecc{x}} R^*(\vecc{k'})d^3k'\rangle=\int d^3k\frac{P(k)}{(2\pi)^3}
=\int\limits_0^{\infty}\frac{dk}{k}\mathcal{P}(k),
\ee
where $\mathcal{P}(k)=k^3 P(k)/(2\pi^2)$ is also called
power spectrum. The quantity $\mathcal{P}(k)$ measures the contribution of a
logarithmic interval of $k$ to the total power of fluctuations. Usually, $\mathcal{P}(k)$ is approximated as follows,
\be
\mathcal{P}_s(k)=A_s\left(\frac{k}{k_*}\right)^{n_s-1}, \label{scalarspect}
\ee
where $k_*$ is some fiducial momentum, $A_s$ is the amplitude of the power
spectrum at this momentum, and $n_s$ is called the spectral tilt. Subscript $s$ refers to scalar perturbations. In 1960's, Zel'dovich and Harrison predicted the flat spectrum of perturbations, that is 
the spectrum with $n_s=1$. 
Approximately flat, Harrison--Zel'dovich spectrum is consistent with the 
 inflationary scenario.

CMB data (and also data on the distribution of
galaxies in space) are obviously sensitive to both the amplitude
$A_s$ and the tilt $n_s$. The amplitude determines the overall magnitude
of the CMB temperature anisotropy, while the tilt gives the relative
power at high and low multipoles. Both parameters can be measured rather
precisely. The current values with $k_*=(500~{\rm Mpc})^{-1}$ are  \cite{WMAPKom}:
\be
&& A_s= (2.46 \pm 0.09)\cdot 10^{-9}, 
\\
&& n_s = 0.960 \pm 0.014.
\ee
Note that the Harrison--Zel'dovich value $n_s=1$ is somewhat disfavored.
The quoted values are obtained, however, under the assumption of the
absence of tensor perturbations (see below)
and exact form (\ref{scalarspect}) of the primordial power
spectrum. Relaxing these assumptions makes the allowed intervals
of $A_s$ and $n_s$ wider. 
Note also that $A_s$ characterizes the perturbation {\it squared};
the amplitude of perturbation itself is of order
\be
\sqrt{A_s} \simeq 5 \cdot 10^{-5}.
\label{may31-3}
\ee
%Of course,
According to the formula (\ref{may31-2}), $\R(k)$ is a relative amplitude of initial perturbations. Also, $\sqrt{A_s}$ measures $\R(k_*)$, so $\sqrt{A_s}$ is approximately equal to the relative CMB temperature fluctuation $\de T/T$.
% it is of the same order as the CMB temperature
%fluctuation $\delta T/T$.

\subsection{Inflation and generation of perturbations}
Inflation is  fast, nearly exponential expansion of the Universe:
\be
a(t)\propto \exp\left(\int Hdt\right), \quad H \approx {\rm const}.
\label{may31-4}
\ee
Inflation occurs if the Universe is filled with a scalar field $\varphi$ 
(inflaton) which has nonvanishing
 scalar potential $V(\varphi)$.
The homogeneous field $\varphi$ satisfies the equation
\be
\ddot{\varphi}+3H\dot{\varphi}=-\frac{d V}{d \varphi}.
\ee  
If $H$ is large (the Universe expands rapidly) and $d V/d \varphi$ is small, the field $\varphi$ varies slowly in time. The Friedmann equation in this case is $H^2=8\pi G V(\varphi)/3$, so 
if $\varphi$ varies slowly, then $V(\varphi)$ and thus $H$ also vary 
slowly. This gives a self-consistent mechanism of inflation. 

The perturbations about the homogeneous field $\varphi$ 
evolve according to the same equation as in Section~\ref{oscsect}
(one can show that the
effect of $V(\varphi)$ on $\delta \varphi$ is negligible):
\be
\ddot{\de\varphi}+3H\dot{\de\varphi}+\frac{k^2}{a^2}\de\varphi=0.
\ee
Recall that if $k/(aH)\gg 1$, the mode is sub-horizon, and if $k/(aH) \ll 1$ the mode is super-horizon. In the
inflationary regime one has
$a(t)\propto \exp(Ht)$, $H$ is almost constant, and $k/(aH)$ decreases
in time. 
% So, in inflationary regime given mode is sub-horizontal at the beginning, and then it becomes super-horizontal. This consequence is reverse with respect to what is observed in standard matter--dominated or radiation--dominated Universe.
 A mode of a fixed $k$ is sub-horizon at the beginning of inflation. Then 
it {\it exits} the horizon, and the mode's evolution terminates. Note that this sequence of events
is reverse as compared to what happens in the
matter--dominated or radiation--dominated Universe.  So, to find 
the perturbations in the Universe towards the end of inflation, 
one needs to calculate the inflaton perturbation 
at the moment of the horizon crossing\footnote{Note that the meaning of the moments denoted by the symbol
$\times$ is different in this Section and in Section~\ref{oscsect}. 
A given mode crosses the horizon at least twice, for the
first time at inflation (it turns from sub-horizon to super-horizon), and for
the second time after inflation (it becomes sub-horizon again). In this 
Section $\times$ denotes the first crossing, and in Section~\ref{oscsect} the second one.} $\de\varphi_{\times}$. There are two quantities of the right dimension: $k/a$ and $H$. But at the time of the horizon exit one has $k/a=H$. So, by dimensional argument, one obtains 
that at the time of the horizon exit
\be
\de\varphi_\times\sim H,
\ee
and $\de\varphi_\times$ is almost independent of $k$.
This value remains intact until the end of inflation. 

After inflation ends, the energy stored in the inflaton field
gets converted into heat, and the Universe enters the hot
Big Bang epoch. The inflaton perturbations get reprocessed
into scalar perturbations (energy density perturbations of
the hot gas of particles and gravitational potentials). These 
are
automatically in the adiabatic mode, since the
perturbations in all forms of matter have one and the same
origin everywhere in space\footnote{In inflationary models with
more than one scalar field, 
the generation of an admixture of entropy modes is also possible.}.
Since the parameters in the inflating Universe are almost time-independent
in the time interval when all interesting modes exit the horizon,
inflation
produces nearly Harrison--Zel'dovich, flat spectrum (we have already noticed
that $\de \varphi_\times$ is almost independent of $k$).
Yet the parameters are not exactly time-independent at inflation,
so the predicted spectral tilt $(n_s-1)$ is small but nonzero.
It can be positive or negative, depending on the shape of the scalar
potential $V(\varphi)$. In particular, it is {\it negative} for the
simplest power-law potentials like
\be
V(\varphi) = \frac{m^2}{2} \varphi^2 \;\;\; {\rm or} \;\;\;
V(\varphi) = \frac{\lambda}{4} \varphi^4. 
\label{may31-6}
\ee
Finally,   the primordial perturbations in the field $\varphi$ have to be 
small in amplitude to generate small amplitude of the scalar perturbations
(\ref{may31-3}).
%Since the potential of the inflaton is flat, its fluctuations are linear.
Thus, $\delta \varphi$ can be treated in the linear approximation. 
Linear quantum fluctuations of scalar fields are known to be Gaussian random variables, so inflation naturally explains this property of density perturbations. We conclude that the inflationary mechanism of the generation of
scalar perturbations is consistent with everything we know about 
them.

\begin{figure}[h!]
\begin{center}
\includegraphics[width=0.7\linewidth]{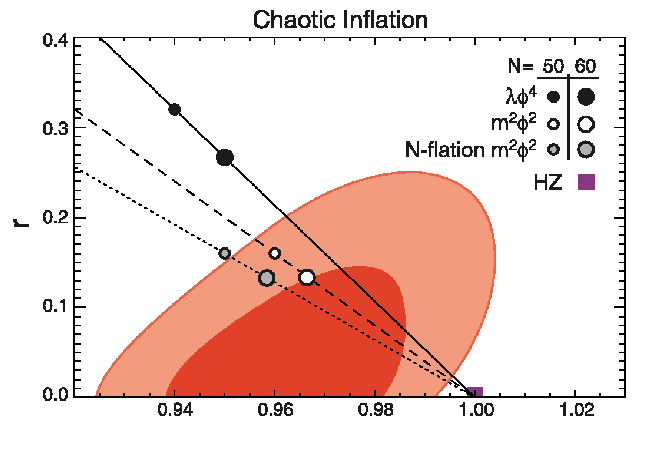}
\caption{~The allowed regions of $n_s$ and $r$  \cite{WMAPKom}. The 
contours show 68\% and 95\% CL 
from WMAP+BAO+SN. The label ``HZ'' corresponds to the flat, Harrison--Zel'dovich scalar spectrum in the absence of tensor perturbations. Also shown are
the predictions of various inflationary models, whose potentials
are given on the right. $N$-flation is a model with numerous
inflaton fields.}
\label{tiltr}
\end{center}
\end{figure}

Actually, the derivation of approximately flat spectrum does not depend on whether we deal with scalar or tensor field. So, 
inflation generates also
{\it tensor perturbations} (transverse traceless perturbations of spatial
metric $h_{ij}$, i. e., gravitational waves). We 
obtain the same picture for them: primordial tensor perturbations
are Gaussian random field with almost flat power
spectrum. 
The standard approximation for primordial 
tensor perturbations spectrum (omitting the details concerning polarization) is:
\be
\mathcal{P}_T(k)=A_T\left(\frac{k}{k_*}\right)^{n_T}.
\ee
Note the flat spectrum corresponds to $n_T=0$, unlike in (\ref{scalarspect}). 
It is convenient to introduce
the parameter $r=\mathcal{P}_T/\mathcal{P}_s$ which
measures the ratio of tensor to scalar perturbations. The simplest theories of inflation like (\ref{may31-6})
predict $r\sim 0.1 - 0.3$. 
The effect of tensor perturbations
has not yet been reliably  observed, but there is great effort to discover it. The allowed regions in the plane $(n_s,r)$ are shown in Fig.\ref{tiltr}.
The detection of tensor perturbations is an extremely
important test for the
simplest inflationary models. The flat primordial
spectrum of scalar perturbations is predicted by some theories other than inflation, but at the moment only inflation predicts the flat spectrum of primordial gravitational waves.

 The tensor perturbations behave at the hot Big Bang epoch exactly in the 
same way as massless scalar field studied in Section~\ref{oscsect}.
These are purely metric perturbations (essentially no perturbations in matter):
\be
ds^2 = dt^2 - a^2(t) [\delta_{ij} - h_{ij}(\vecc{x},t)]dx^i dx^j,
\ee
where $h_{ij}$ is a transverse traceless tensor,
$\partial_i h_{ij} = 0; h_{ii}=0$. Before the horizon re-entry,
metric perturbation $h_{ij}$ stays constant in time; after the horizon
entry it oscillates and its amplitude decreases as $1/a$. For modes that
cross the horizon at radiation domination one has
(compare with (\ref{may31-7}); we omit indices $i,j$)
\be  
h(\vecc{k},t)=h_{(i)}(\vecc{k})\cdot
\frac{a(t_{\times})}{a(t)}\cdot\sin\left(k\int\limits_0^t~\frac{dt'}{a(t')}\right),
\label{may31-10}
\ee
where $h_{(i)}$ is the primordial amplitude (Gaussian random field).
The phase of oscillations is different for modes that cross the
horizon at matter domination.

\begin{figure}[h!]
\begin{center}
\includegraphics[width=0.6\linewidth]{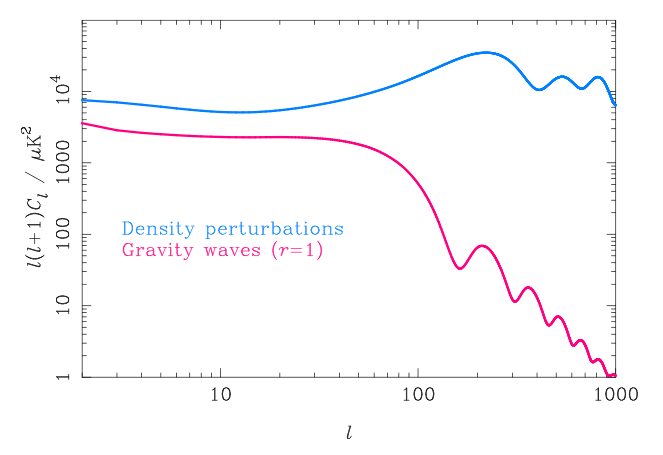}
\caption{~The effect of gravity waves on CMB  \cite{Challinor}.}
\label{cmbgravit}
\end{center}
\end{figure}

For nearly flat primordial power spectrum,
the shorter the mode, the earlier it crosses the horizon,
the smaller $a(t_\times)$, the lower the present amplitude.
For this reason it is 
almost hopeless to detect relic gravity waves 
directly by existing and planned detectors of gravitational waves.
Presently, the best constraints on tensor perturbations
come from the analysis of the CMB temperature anisotropy.
The effect of gravitational waves on the CMB temperature anisotropy 
is given by:
\be
\frac{\de T}{T}=\int\limits_{t_r}^{t_0}~dt~\dot{h}_{ij}n^in^j,
\ee
where $n^i$ is the unit vector along the photon path
and the integration is performed along the photon world line. 
This is nothing but the integrated Sachs--Wolfe effect;
other effects we considered in Section~\ref{SW-D-ISW}
are absent for tensor perturbations.  The modes which 
are super-horizon at  
recombination ($l\lesssim 50$) have not changed since inflation, as they have never been sub-horizon after inflation
but before recombination. So, their spectrum at recombination 
is flat, and they generate flat contribution
to the CMB temperature angular spectrum at low $l$. At higher multipoles,
shorter gravity waves contribute (recall the relationship
(\ref{may30-2})), their amplitudes decrease with $k$, and 
hence their effect on the CMB temperature decreases with $l$.
This is shown in
Fig.\ref{cmbgravit}. The oscillations in the angular spectrum
are due to the oscillatory behavior in (\ref{may31-10}).

Clearly, it is hard to disentangle the contributions of scalar and tensor
perturbations to the CMB temperature anisotropy. In particular,
tensor perturbations and red scalar tilt ($n_s<1$) have similar 
effect of the enhancement of low multipoles (this is the basic
reason for the shape of the allowed region in Fig.~\ref{tiltr}).
Potentially more powerful way to detect primordial gravity waves
is the measurement of the CMB polarization.

\section{CMB polarization and primordial gravitational waves}
\subsection{The physical mechanism of polarization\label{Polmechanism}}
%The map of CMB polarization is presented in Fig.\ref{cmbpole}

CMB polarization is generated by 
a simple mechanism: light gets polarized when it 
scatters off free electrons. 
Indeed, when the light is scattered off an electric charge, 
the electromagnetic wave induces the oscillations of the
charge in the direction of electric field. Electric field is 
perpendicular to the direction of the
wave propagation. The oscillating charge emits 
radiation only in directions perpendicular to the direction of 
its oscillations. So, if the initial light is polarized 
in the plane 
formed by the direction of initial propagation and direction of 
scattered light (scattering plane), smaller amount of radiation is emitted
as compared with incoming light polarized normal to the 
scattering plane. 
%This is a very general property of  the Thomson scattering.
 A photon with polarization 
normal to the scattering plane scatters at the cross-section
\be
\frac{d\sigma}{d\Omega}=\frac{3\sigma_T}{8\pi},
\ee
and its polarization remains unchanged. If the initial
polarization is in the scattering plane, it remains in that plane,
and
the cross-section is
\be
\frac{d\sigma}{d\Omega}=\frac{3\sigma_T}{8\pi}\cos^2\theta,
\ee
where $\theta$ is the scattering angle. 
If at some position the flux of photons is anisotropic, the 
radiation scattered there towards an observer
is linearly polarized. The local anisotropy 
before the last scattering can be caused by both temperature inhomogeneities 
and gravitational waves. The effect involves the factor 
$\lambda_\gamma/\lambda(k)$, where $\lambda_\gamma$ is the 
photon mean free path just before last scattering, 
and $\lambda(k)$ is the physical wavelength of the perturbation which 
is responsible  for the local
anisotropy. So, the estimate for the degree of polarization $P$ induced by temperature inhomogeneities is
\be
P=\frac{\lambda_\gamma}{\lambda(k)}\frac{\de T}{T}.
\ee
For $l\sim 100$ (this 
roughly corresponds to the horizon at recombination) the degree 
of polarization is approximately $10^{-6}$.
\subsection{$E-$ and $B-$modes}
Consider a polarimeter 
oriented along a vector $\vecc{n}$ normal 
to the line of sight. The  intensity it measures is
given by
\be
I_{\vecc{n}}=\langle(\vecc{E}\vecc{n})^2\rangle=n_an_b\langle E_aE_b\rangle= \langle E^2 \rangle P_{ab} n_a n_b,
\ee
where the polarization tensor is defined as follows:
\be
P_{ab}=\frac{\langle E_aE_b \rangle}{\langle E^2\rangle }.
\ee
All averages here are the time
averages over the period of the wave. For linear polarization, the 
tensor $P_{ab}$ is real. Additionally, it is symmetric and satisfies the constraint ${\rm Tr} (P)=P_{aa}=1$. A symmetric $2$-tensor 
has three parameters, and since 
the polarization tensor
satisfies one additional trace
constraint, it is characterized by 
only two parameters. The eigenvectors of this tensor are the directions of maximal and minimal intensity. As is clear from the above discussion, the direction of maximal intensity near a
hot spot is perpendicular to the direction towards the spot. Near a 
cold spot the direction of maximal intensity 
is the direction towards the spot.

\begin{figure}[h!]
\begin{center}
\includegraphics[width=0.7\linewidth]{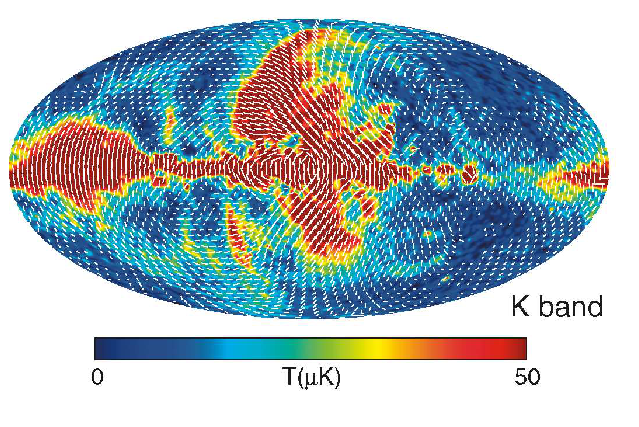}
\caption{~CMB polarization map in WMAP K band.\cite{WMAP}}
\label{cmbpole}
\end{center}
\end{figure}

CMB polarization can be visualized by showing the direction of
maximal polarization and the degree of polarization
${\mathcal P}= \sqrt{1-4{\rm det}P}$. The actual
CMB polarization map is shown in Fig.\ref{cmbpole}.

\begin{figure}[h!]
\begin{center}
\includegraphics[width=1.0\linewidth]{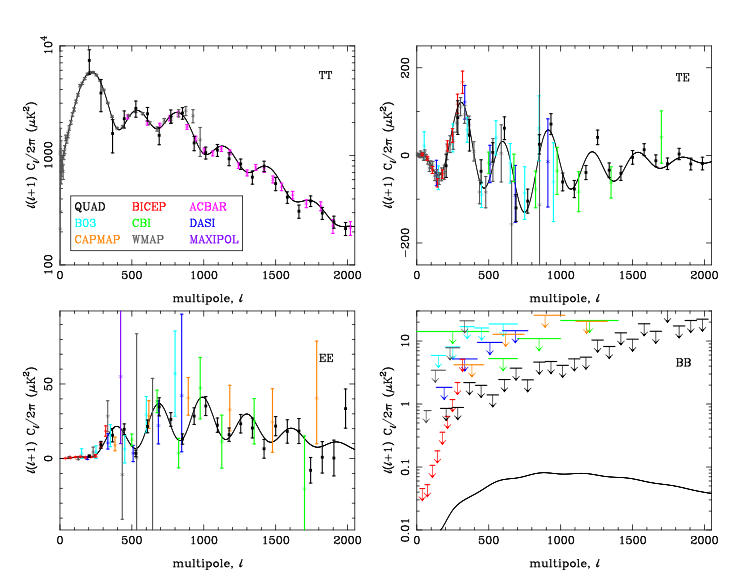}
\caption{Correlators of various 
polarization potentials and 
temperature as functions of $l$  \cite{CMBPol}. 
$C^{EE}$ and $C_l^{BB}$ are multiplied by $T_0^2$,
and $C_l^{TE}$ is multiplied by $T_0$, so all these quantities 
have the same dimension as $C^{TT}$.
The $BB$-mode has not yet been detected; 
the lower right panel 
shows the existing limits and predicted spectrum generated via
gravitational lensing by structures in the late Universe in a
model without primordial
gravitational waves.}
\label{cmbeebb}
\end{center}
\end{figure}

It is convenient to
parametrize the polarization tensor  as follows:
\be
 P_{ab}-\frac12\de_{ab}=-\{\nabla_a\nabla_b\} P_E-\{\eps^c_a\nabla_b\nabla_c\} P_{B},
\ee
where $\{A_{ab}\}=(A_{ab}+A_{ba}-\de_{ab}A_{cc})/2$. The modes
 generated by the potentials $P_E$ and $P_B$
are called $E$-mode and $B$-mode, respectively, in distant
analogy with electric and magnetic fields. The $B$-mode 
is obviously parity-odd, while the $E$-mode is parity-even. 
{\it Scalar perturbations  generate only $E$-mode}\footnote{Some
admixture of $B$-mode is generated at small angular scales
from $E$-mode via gravitational
lensing by structures (clusters of galaxies, etc.) in the late Universe.}, 
{\it while tensor perturbations generate both modes}. This is because that parity-odd B-mode can noy generate parity-even scalar perturbations.  Similarly 
to the expansion of temperature, one expands potentials 
of CMB polarization in spherical harmonics; omitting $l$-dependent
factors, we write
\be 
P_{E,B}(\vecc{n})=\sum_{l,m}a_{lm}^{E,B}Y_{lm}(\vecc{n}).
\ee
Equivalently, $a_{lm}^{E,B}$ are the expansion coefficients of $P_{ab}$ in
parity-even and parity-odd tensor  spherical harmonics.
Polarization observables are the correlators
\be
C^{EE}(l)\propto \; \langle a^{E}_{lm}a^E_{lm}\rangle, \\
C^{BB}(l)\propto \; \langle a^{B}_{lm}a^B_{lm} \rangle.
\ee
Another possible correlator, $\langle a_{lm}^Ea_{lm}^B\rangle$, vanishes
because of its odd parity ($a^E$ are parity-even and $a^B$ are parity-odd).  Since 
CMB polarization is caused predominantly by local
anisotropies in the medium, which are generated by the temperature
inhomogeneities, 
one naturally expects that polarization will correlate with the CMB
temperature fluctuations. Hence, one introduces the cross-correlator
\be
C^{TE}(l)\propto \; \langle a^E_{lm}a^T_{lm} \rangle.
\ee
The cross-correlator $C^{TB}(l)\propto \; \langle a^B_{lm}a^T_{lm} \rangle$  also vanishes because
of the parity argument. 
In observations, only $C^{TE}$ and $C^{EE}$ correlators were discovered
so far, and there exist upper limits on $C^{BB}$, see Fig.\ref{cmbeebb}.

\begin{figure}[h!]
\begin{center}
\includegraphics[width=0.4\linewidth]{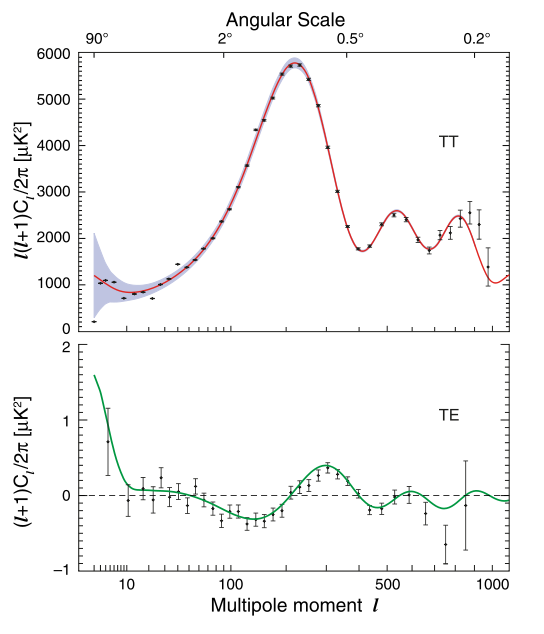}
\caption{~Effects of scalar perturbations:
$C^{TE}_l$ compared with $C^{TT}_l$ \cite{WMAP}. 
Zeroes of $C^{TE}_l$ approximately coincide 
with maxima and minima of $C^{TT}_l$ and vice versa.}
\label{cmbttte}
\end{center}
\end{figure}

 Scalar perturbations contribute to 
polarization mainly because of the motion of the baryon--photon medium. 
Photons coming along the direction of motion are hotter than photons coming from the opposite direction. This is precisely the Doppler effect that produces local anisotropy before the very last scattering, and hence CMB polarization. 
It is clear from the discussion in Section~\ref{Polmechanism} that this 
local anisotropy by itself is insufficient for producing CMB polarization
(effects due to photons coming from opposite directions would cancel out),
so there is still a suppression factor $\lambda_\gamma/\lambda(k)$.
For the same reason,
the effect of density inhomogeneities on CMB
polarization is of the second order 
in $\lambda_\gamma/\lambda(k)$.
% So,  zeroes of $C_l^{EE}$ 
%are zeroes of velocity. 
In acoustic waves, 
the phase of oscillations of
velocity is shifted by $\pi/2$ from the phase of
oscillations of  density. The effect of this shift
is that the oscillations of $C_l^{TE}$ are shifted by half-period
with respect to oscillations in $C_l^{TT}$ (recall that the 
oscillating part of the CMB temperature
anisotropy spectrum $C^{TT}_l$ comes mainly
from $\delta \rho_\gamma$, recall also the correspondence
(\ref{may30-2}); in the previous Sections
we did not write the superscript $TT$ in $C_l$): 
zeroes of $C^{TE}_l$ should coincide with maxima and minima 
of $C^{TT}_l$ .
This can be seen in Fig.\ref{cmbttte}. 
% Usually, $C^{EE}$ is multiplied by $T_0^2$, 
%and then $C^{EE}$ has the same dimension as $C^{TT}$.

The effect of gravitational waves on CMB is much more interesting. 
At relatively low multipoles, primordial gravitational waves (tensor
perturbations) are the only possible source of considerable
$B$-mode of CMB polarization. Thus, the detection of the $B$-mode
will be the detection of primordial gravity waves. Measuring their
power spectrum will be of paramount importance for understanding
the origin of cosmological perturbations.

\begin{figure}[h!]
\begin{center}
\includegraphics[width=0.6\linewidth]{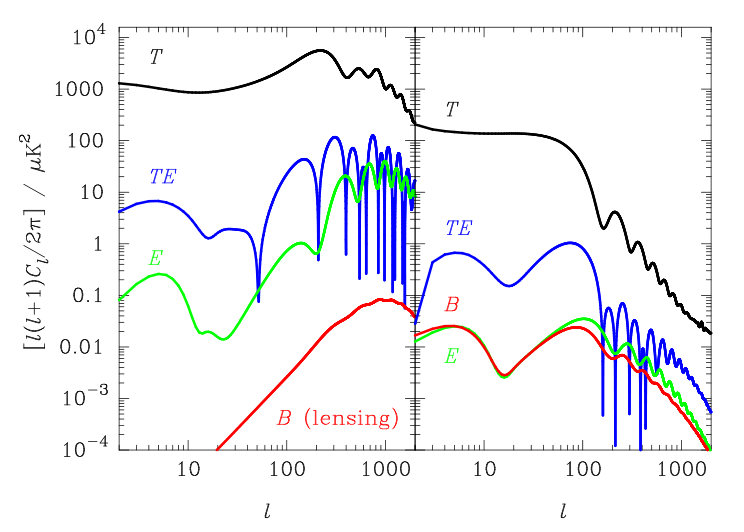}
\caption{~The effect of scalar perturbations
(left panel) and gravity waves (right panel) on CMB  \cite{Challinorconst}.
Primordial scalar and tensor spectra are assumed to be flat.
The relative normalization is arbitrary.}
\label{cmbpolegrav}
\end{center}
\end{figure}

As we have already mentioned, the smaller the wavelength of a gravitational 
wave, the earlier it becomes sub-horizon and the smaller its amplitude
at recombination. So, the effect of gravitational waves is small 
for small wavelengths, and hence for large $l$. On the other hand, the effect of gravitational waves 
at recombination contains the factor $(\lambda_\gamma/\lambda(k))^2$, so it is 
small for very large wavelengths and thus for very
small $l$. The maximal effect is at $l\sim 50$. We note, however, 
that the interaction of gravitational waves with light, which leads to 
CMB polarization, occurs not only at the time of last scattering. Approximately at $z\sim 10$, 
intergalactic gas was reionized by the light of first very massive stars with $M\sim 100~M_{\odot}$. Tensor
modes of wavelengths  approximately equal to the size of the horizon at this
 reionization epoch are  most important in this situation. 
This means that the effect is most pronounced at $l\lesssim 10$. This is 
seen in Fig.~\ref{cmbpolegrav}, a wide peak  at  $l\sim 2-7$ is 
precisely due to reionization.  The optical depth at 
reionization (i. e., the probability for a photon to 
be scattered) is about $0.1$, 
this is the reason for the smallness of the effect.

Special experiments intended to measure the CMB polarization, such as CMBPole, are being prepared. 
The sensitivity of CMBPole to $r=A_T/A_s$ is approximately $10\%$, and it is planned to increase the sensitivity to the level of $1\%$ in the next
generation experiments.

\section{Summary}

To summarize, measurements of the CMB temperature anisotropy and polarization is a powerful tool in cosmology. Combined with other methods, they are capable of determining with high precision the cosmological parameters characterizing the
recent Universe and revealing the properties of the
primordial perturbations. The current picture of the Universe is quite
simple. As to the recent Universe, the data are consistent with
the spatially flat $\Lambda$CDM model whose ingredients are
time-independent dark energy density,
cold dark matter, baryons, electrons, photons and fairly light
neutrinos ($m_\nu < 0.2$~eV for each of the neutrino species).
Known properties of scalar perturbations are also simple: they were
generated before the hot Big Bang epoch, have no decaying
super-horizon modes, are adiabatic, Gaussian and have flat or
nearly flat power spectrum. Tensor perturbations have not
been discovered, though the current bound on their amplitude is not
particularly strong. All these properties are consistent with
the inflationary mechanism of the generation of cosmological perturbations,
but inflation is not the only option for the moment.

What can one expect to be discovered in future? On the recent Universe 
side, the major issues are whether the dark energy density is exactly
constant in time or not, and what are dark matter particles.
CMB alone will not shed much light on these issues, but its analysis will
be instrumental in combination with other methods. 
Concerning the cosmological perturbations, one intrigue is whether
the scalar power spectrum is exactly flat or not. Even more interesting
are tensor perturbations whose primordial amplitude is predicted
by the simplest inflationary models to be quite large; furthermore,
the smoking-gun prediction of inflation is nearly flat tensor spectrum.
Unexpected discoveries cannot be excluded either, like sizeable
non-Gaussianity or admixture of entropy modes in scalar
perturbations. All these are the tasks for future CMB experiments,
which will thus serve as windows to the extremely early
cosmological epoch preceding the hot Big Bang stage.

The work of V.R. has been supported in part by Russian Foundation for Basic Research grant 08-02-00473. The work of A. V. is supported in part by Russian Foundation of Basic Research grant 090200393 and by Federal Agency for Science and Innovations of Russian federation under contract 02.740.11.5194, by Federal Programm ``Scientific and pedagogical specialists of innovation Russia'', contract number 02.740.11.0250 and by Dynasty foundation.


\begin{thebibliography}{12}

% [1]
\bibitem{WMAP} G.~Hinshaw {\it et al.},
  Astrophys.\ J.\ Suppl.\  {\bf 180}, 225 (2009); arXiv:0803.0732 [astro-ph].

% [2]
\bibitem{Peacock} J.~A.~Peacock,
{\it Cosmological Physics} (Cambridge University Press, 1999).

%[3]~~
\bibitem{Dodelson}
 S.~Dodelson, {\it Modern Cosmology} (Academic Press, Amsterdam, 2003).

% [4]~~ 
\bibitem{Mukhanov} V.~Mukhanov, {\it Physical Foundations of Cosmology} (Cambridge University Press, 2005).

%[5]~~
\bibitem{Naselsky}  P.~Naselsky, D.~Novikov, and I.~Novikov, {\it The Physics of Cosmic
 Microwave Background} (Cambridge University Press, 2006).

%[6]~~ 
\bibitem{Weinberg}  S.~Weinberg,
{\it Cosmology} (Oxford University Press, 2008).

%[7]~~
\bibitem{Giovannini} M.~Giovannini, {\it A Primer on the Physics of the Cosmic Microwave Background} (World Sci., Singapore, 2008).

%[8]~~ 
\bibitem{Durrer} R.~Durrer, {\it Cosmic Microwave Background} (Cambridge University Press, 2008).

%[9]~~ 
\bibitem{GorbunovRubakov}  D.~S.~Gorbunov, V.~A.~Rubakov,
{\it Introduction to the Theory of the Early Universe. 
Cosmological Perturbations. Inflationary Theory 
 }
% (in Russian) (URSS, Moscow, 2009) (
English translation to appear, World Scientific, Singapore.
%).

%[10]~~ 
\bibitem{Kosowsky}  A.~Kosowsky, Annals Phys. {\bf 246}, 49 (1996); astro-ph/9501045.

%[11]
\bibitem{Challinor} A.~Challinor, Lect. Notes Phys. {\bf 653}, 71 (2004); astro-ph/0403344.

%[12]~~ 
\bibitem{ACBAR}  C.~L.~Reichardt {\it et al.}, Astrophys. J. {\bf 694}, 1200 (2009); arXiv:0801.1491 [astro-ph].

%[13]~~ 
\bibitem{strings}
J.~Urrestilla, N.~Bevis, M.~Hindmarsh, M.~Kunz, and A.~R.~Liddle, JCAP {\bf 0807}, 010 (2008); arXiv:0711.1842 [astro-ph].

%[14]~~ 
\bibitem{WMAPKom} E.~Komatsu {\it et al.}, Astrophys. J. Suppl. {\bf 180}, 330 (2009); arXiv:0803.0547 [astro-ph].

%[15]~~
\bibitem{BAO} 
W.~J.~Percival {\it et al.}, Mon. Not. Roy. Astron. Soc. {\bf 381}, 1053 (2007); arXiv:0705.3323 [astro-ph].

%[16]~~ 
\bibitem{BAOPeak}
 D.~J.~Eisenstein {\it et al.}, Astrophys. J. {\bf 633}, 560 (2005); astro-ph/0501171.

%[17]~~
\bibitem{CMBPol}  
M.~L.~Brown {\it et al.}, Astrophys. J. {\bf 705}, 978 (2009); arXiv:0906.1003 [astro-ph.CO].

%[18]~~ 
\bibitem{Challinorconst} A.~Challinor, astro-ph/0606548.

\end{thebibliography}
\end{document}